# The Compressive and rarefactive dust acoustic solitary waves with two different temperatures for both electrons and ions

Rasool Barazandeh Kian [1], Mohammad Hossein Mahdieh [2]

*Department of Physics, University of Science & Technology, Narmak, Tehran, Iran*

The nonlinear propagation of dust acoustic solitary waves (DASWs) in an unmagnetized dusty plasma consisting of negatively charged dust fluid, Boltzmann distributed electrons with two different temperatures, Boltzmann distributed ions with two distinct temperatures are investigated. By employing the reductive perturbation technique that is valid for a small but finite amplitude limit, the Korteweg–de-Vries (K-dV) equation, have been derived. Our results reveal that the main quantities of DASWs (such as amplitude and width) are affected by two different temperatures electrons (as well as ions), temperature ratios, and the number densities of two species of ions. It is shown that both positive and negative potential DASWs occur in this case.

## INTRODUCTION

A dusty plasma which consisting of electrons, ions, charged dust particles is found in space environments, such as earth's ionosphere, planetary rings, interstellar medium, Solar nebula, cometary tails [1-7], in the laboratory devices such as dc and RF discharges machines, plasma processing reactors, fusion plasma devices, and in the industrial plasma [8-14].

In these devices, the particles are either formed as a result of agglomeration of reactive species or deliberately injected into the plasma [15-17]. The presence of dust component in a plasma can significantly modify the existing ordinary modes [viz., ion acoustic (IA) waves] and introduce new nonlinear waves at different dynamical scales. One of these modes is low frequency dust acoustic (DA) wave (with a phase speed lying in between the ion and dust thermal velocities), which was first predicted theoretically by Rao *et al.* [18]. In laboratory the spectra of DA waves have been experimentally observed by Barkan et al. [19].

The nonlinear DA waves, have received a great deal of research interest in understanding the fundamental properties of localized electrostatic structures in space and laboratory dusty plasmas [20-27]. It is well known that two temperature electrons are found to occur both in space and laboratory environment [28-31]. On the other hand, in laboratory and space plasmas, it is most evident that the ions are usually in different temperatures in a large plasma column. Therefore, in many plasmas, electrons and ions grouped into two distinct components known as low and high temperature electrons as well as low and high temperature ions. Several researchers have worked with two temperature electron populations in a dusty plasma to study the solitary structures [32-39]. And a good number of researchers have studied the properties of DASWs in a dusty plasma consisting of two temperature ions [40-52]. Kerong et al. [53] investigated the effect of polarization and charge gradient forces on propagation of DASWs in an unmagnetized dusty plasma. They found that polarization force and charge gradient force have opposite effects on the amplitude and width of DASWs. Mamun et al. [54] investigated DA wave in an opposite polarity dusty plasma. They showed that the amplitude of DA wave electrostatic structures decreases with the increase in the positive dust number density. Kavitha et al. [55] studied the propagation of electrostatic DASWs in an unmagnetized dusty plasma consisting positively and negatively charged mobile dust with the Boltzmann distributed electrons and nonthermal ions. They found that existence of nonthermal ions modify the nature of the electrostatic DASWs structure. Rezaiinia et al. [56] investigated the nonlinear propagation of DASWs in a magnetized dusty plasma consisting of Boltzmann distributed electrons and ions with two distinct temperatures following a nonextensive distribution. They have applied reductive perturbation method (which is appropriate for small finite amplitude limit) to drive K-dV equation. They found that rarefactive and compressive potential DASWs can be formed in such dusty plasma.

---

[1] *rasool_barazandeh@physics.iust.ac.ir*
[2] *Corresponding author: mahdm@iust.ac.ir*

Malik et al. [57] studied DASWs in a magnetized dusty plasma considering two different temperature ions and Boltzmann distributed electrons. In their study, they found that the effect of hot ions to be significant when their number is very high as compared to the number of cold ions. Also they found that the presence of more electrons in the dusty plasma increases the amplitude of the DASWs. Emamuddin et al. [58] theoretically investigated the propagation of DA waves in a magnetized dusty plasma consisting of two distinct temperature electrons, nonthermal ions and negatively charged mobile dust grains. Their analysis show that the characteristics and the properties of the DASWs are significantly modified by the presence of external magnetic field, temperature ratio of ions, relative number densities of ions as well as electrons. Shamy et al. [59] studied DASWs for dusty plasmas consisting of hot dust fluid, nonisothermal ions and two different temperature electrons. They derived K-dV equation and modified K-dV equation at critical ion density. They have shown that the presence of a second component of electrons modifies the nature of DA solitary waves structures. The above-mentioned studies of DA solitary waves in a magnetized or an unmagnetized dusty plasma with two different temperature electrons or two different temperature ions, show a good characteristic dependency of these nonlinear structures on the second temperature of electrons as well as ions in dusty plasma which motivate us to study DASWs in a dusty plasma with two different temperatures for both electrons and ions.

## BASIC EQUATION

We consider a five-component plasma, which consists of negatively charged dust grains, two types of isothermal electrons with temperatures $T_{el}$ and $T_{eh}$, two types of isothermal ions with temperatures $T_{il}$ and $T_{ih}$. The charge neutrality at equilibrium requires that,

$$n_{il0} + n_{ih0} = n_{el0} + n_{eh0} + n_{d0}Z_d \tag{1}$$

where $n_{d0}$ is unperturbed dust number density, $n_{el0}$ and $n_{eh0}$ are equilibrium electron number densities at low ($T_{el}$) and high ($T_{eh}$) temperature electrons respectively. $n_{il0}$, $n_{ih0}$ are the unperturbed ion number densities at low ($T_{il}$) and high ($T_{ih}$) temperature ions respectively. $Z_d$ is the number of charge on the dust grains. From equation (1) we have

$$\alpha = \frac{n_{do}Z_d}{n_{il0}} = 1 - \delta_1 - \delta_2 + \delta_3 \tag{2}$$

Where $\delta_1 = \frac{n_{elo}}{n_{ilo}}$, $\delta_2 = \frac{n_{eho}}{n_{ilo}}$, $\delta_3 = \frac{n_{iho}}{n_{ilo}}$

First we define an effective temperature being similar to that in reference [49].

$$T_{eff} = \frac{\alpha T_{el} T_{eh} T_{il} T_{ih}}{\delta_1 T_{eh} T_{il} T_{ih} + \delta_2 T_{el} T_{il} T_{ih} + T_{el} T_{eh} T_{ih} + \delta_3 T_{el} T_{eh} T_{il}} \quad (3)$$

According to one dimensional propagation, the dynamics of dust particles are governed by normalized equations of fluid equations:

$$\frac{\partial n_d}{\partial t} + \frac{\partial (n_d u_d)}{\partial x} = 0 \quad (4)$$

$$\frac{\partial u_d}{\partial t} + u_d \frac{\partial u_d}{\partial x} = \frac{\partial \phi}{\partial x} \quad (5)$$

$$\frac{\partial^2 \phi}{\partial x^2} = n_d + \frac{\delta_1}{\alpha} e^{s\theta_1 \phi} + \frac{\delta_2}{\alpha} e^{s\theta_2 \phi} - \frac{1}{\alpha} e^{-s\phi} - \frac{\delta_3}{\alpha} e^{-s\theta_3 \phi} \quad (6)$$

Where $\frac{T_{eff}}{T_{il}} = s = \frac{\alpha}{1 + \delta_1 \theta_1 + \delta_2 \theta_2 + \delta_3 \theta_3}$, $\theta_1 = \frac{T_{il}}{T_{el}}$, $\theta_2 = \frac{T_{il}}{T_{eh}}$, $\theta_3 = \frac{T_{il}}{T_{ih}}$. $n_d$ is the dust particle number density normalized to $n_{d0}$, $u_d$ is the dust fluid velocity in $x$ direction normalized to the dust acoustic speed $C_d = (\frac{Z_d T_{eff}}{m_d})^{\frac{1}{2}}$, where $m_d$ is dust mass. $\phi$ is the electrostatic potential normalized to $\frac{T_{eff}}{e}$ with $e$ being the magnitude of the electron charge. The time $t$ and space $x$ variables are normalized by the dust plasma period $\omega_{pd}^{-1} = \sqrt{\frac{m_d}{4\pi n_{do} Z_d^2 e^2}}$ and the Debye length $\lambda_{Dd} = \sqrt{\frac{T_{eff}}{4\pi n_{do} Z_d e^2}}$ respectively.

In order to derive the K-dV equation, we employ the reductive perturbation technique [60] and introduce stretched coordinates as follows:

$$\xi = \varepsilon^{\frac{1}{2}} (x - v_0 t) \quad , \quad \tau = \varepsilon^{\frac{3}{2}} t \quad (7)$$

Where $\varepsilon$ is a small parameter characterizing the strength of nonlinearity ($0 < \varepsilon < 1$) and $v_0$ is the nonlinear wave phase velocity which normalized by $C_d$.

The variables $n_d$, $u_d$, $\phi$ can be expanded in terms of power series of $\varepsilon$:

$$n_d = 1 + \varepsilon n_{d1} + \varepsilon^2 n_{d2} + \dots \quad (8)$$

$$u_d = \varepsilon u_{d1} + \varepsilon^2 u_{d2} + \dots \quad (9)$$

$$\phi = \varepsilon \phi_1 + \varepsilon^2 \phi_2 + \dots \quad (10)$$

Substituting equations (8)-(10) along with the stretching coordinates (7) into equations (4)-(6) and collecting the terms in the different powers of $\varepsilon$, to lowest order in $\varepsilon$, we obtain the following relations

$$n_{d1} = \frac{1}{v_0} u_{d1} \quad (11)$$

$$u_{d1} = -\frac{1}{v_0}\phi_1 \tag{12}$$

$$n_{d1} = -\phi_1 \tag{13}$$

For the next higher order, we obtain

$$v_0^2 \frac{\partial n_{d2}}{\partial \xi} = v_0 \frac{\partial n_{d1}}{\partial \tau} + v_0 \frac{\partial (n_{d1}u_{d1})}{\partial \xi} + v_0 \frac{\partial u_{d2}}{\partial \xi} \tag{14}$$

$$v_0 \frac{\partial u_{d2}}{\partial \xi} = \frac{\partial u_{d1}}{\partial \tau} + u_{d1}\frac{\partial u_{d1}}{\partial \xi} - \frac{\partial \phi_2}{\partial \xi} \tag{15}$$

$$\frac{\partial^2 \phi_1}{\partial \xi^2} = n_{d2} + \phi_2 - \frac{1+\delta_3-\delta_1-\delta_2}{2(1+\delta_1\theta_1+\delta_2\theta_2+\delta_3\theta_3)^2}(1+\delta_3\theta_3^2-\delta_1\theta_1^2-\delta_2\theta_2^2)\phi_1^2 \tag{16}$$

Taking the derivative of equation (16) with respect to $\xi$, and using equations (11)-(15), we eliminate all the second order terms, and finally obtain the K-dV equation

$$\frac{\partial \phi_1}{\partial \tau} + A\phi_1 \frac{\partial \phi_1}{\partial \xi} + B\frac{\partial^3 \phi_1}{\partial \xi^3} = 0 \tag{17}$$

where the nonlinear coefficient $A$ and the dispersion coefficient $B$ are given by

$$A = \frac{1}{2}[\frac{1+\delta_3-\delta_1-\delta_2}{(1+\delta_1\theta_1+\delta_2\theta_2+\delta_3\theta_3)^2}(1+\delta_3\theta_3^2-\delta_1\theta_1^2-\delta_2\theta_2^2)-3] \tag{18}$$

$$B = \frac{1}{2} \tag{19}$$

Equation (17) represents the well-known K-dV equation describing the nonlinear propagation of the DASWs in an unmagnetized dusty plasma with two different temperatures for both electrons and ions.

The stationary soliton-like solution of the K-dV equation can be obtained by transforming the independent variables $\xi$ and $\tau$ to $\chi = \xi - u_0\tau$ and $\tau = \tau$, where $u_0$ is a constant velocity normalized by $C_d$, and imposing the appropriate boundary conditions, namely $\phi \to 0$, $\frac{d\phi}{d\chi} \to 0$, and $\frac{d^2\phi}{d^2\chi} \to 0$, as $\chi \to \pm\infty$. Accordingly, the stationary solitary wave solution of this nonlinear equation is

$$\phi_1 = \phi_{m1} \sec^2 h(\frac{\chi}{\sigma}) \tag{20}$$

where the amplitude $\phi_{m1}$ and the width $\sigma$ are given by $\phi_{m1} = \frac{3u_0}{A}$ and $\sigma = \sqrt{\frac{2}{u_0}}$, respectively.

## RESULTS AND DISCUSSIONS

Equation (20) clearly indicates that both positive and negative solitons exist. The solution (20) also stands for $n_{d1}$ if we replace $\phi_{m1}$ by $-\phi_{m1}$. The coefficient of the nonlinear term $A$ which determines the polarity of solitary waves, depends on equilibrium density ratio parameters $\delta_1, \delta_2, \delta_3$ and temperature ratio parameters $\theta_1, \theta_2, \theta_3$. To preserve the physical meanings, the possible values of these parameters are chosen as $T_{il} < T_{ih} < T_{el} < T_{eh}$ and $n_{el0}, n_{eh0} < n_{il0}$. As $u_0 > 0$, for $A > 0$, positive potential or rarefactive solitary waves (DASWs with density dip) are formed and for $A < 0$, negative potential or compressive solitary waves (DASWs with density hump) are formed. It is clear that an increase in $u_0$ increases the amplitude of the soliton wave, and decreases its width.

For $\delta_1, \delta_2 \simeq 0$, i.e., electron depleted plasma, the values of nonlinear coefficient $A$ and dispersion coefficient $B$ are in good agreement with the findings of [46, 61]. Furthermore, by considering electrons at one temperature ($\vartheta_2 \simeq 0$), the nonlinear coefficient $A$ and dispersion coefficient $B$ are similar to that obtained in Refs. [49-52]. Figs. 1-2 show the variation of the amplitude of the DASWs versus position coordinate $\xi$ and time $\tau$. Fig. 1 shows, small amplitude solitary waves with negative potential (compressive solitary wave) is generated, while Fig. 2 shows, small amplitude solitons with positive potential (rarefactive solitary wave) is formed. The results in these figures show that the amplitude of the soliton is constant with increasing in time $\tau$. Fig. 3 indicates that compressive (Fig. 3a) and rarefactive (Fig. 3b) solitary waves are formed in different values of $u_0$. Obviously, from this figure, one can see that the amplitude (width) of both compressive and rarefactive soliton increases (decreases) as $u_0$ increases.

Fig. 4 which plotted $\phi_1$ vs $\chi$ depicts that there is a range of values of $\delta_3$ (the ratio of density of high temperature ions to density of low temperature ions) in which either compressive or rarefactive solitary waves coexist. It is clear that if $\delta_3 = 6$, compressive solitary waves will be generated in the dusty plasma, and for $\delta_3 = 7$ we will have rarefactive solitary wave in dusty plasma. Fig. 5 depicts the soliton profile for two values of $\theta_3$ (the ratio of low temperature ions to high temperature ions). It is evident that the polarity of DA solitons (compressive or rarefactive) depends sensitively on the $\theta_3$. We see that compressive DA soliton can exist for $\theta_3 = 0.2$ ($T_{eh} = 2T_{el} = 20T_{ih} = 100T_{il}$), while $\theta_3 = 0.05$ ($T_{eh} = 2T_{el} = 5T_{ih} = 100T_{il}$) leads to appearance of rarefactive DA soliton.

Figs. 6-7 show the variation of the amplitudes of the DASWs with the temperature ratios $\theta_1, \theta_2$ and $\theta_3$ for some specified values of ratio densities $\delta_1, \delta_2, \delta_3$ and constant speed of DASWs. These figures show that compressive and rarefactive DASWs may exist in our two temperature electrons and two temperature ions dusty plasma model.

The effect of the temperature of two electron species on the amplitude of DASWs is presented in Fig. 6. In Fig. 6(a) the amplitude of compressive DASWs decreases with increasing the values of both $\theta_1$ and $\theta_2$. While in Fig. 6(b), polarity is changed (amplitude is positive) and rarefactive solitons are formed. It can be seen that the amplitude of soliton increases with increase in $\theta_1$ and $\theta_2$. It must be noted that the change of amplitude by $\theta_1$ (due to the lower temperature electron) is greater than the amplitude change by $\theta_2$ (due to the higher temperature electron), which completely agrees with the result of Mamun et al. [35]. To see the effects of ratio ion temperature on amplitude of DASWs, electrostatic potential $\phi_1$ is plotted against $\chi$ for different values of $\theta_3$ (the ratio of low temperature ion to high temperature ion). In Fig. 7(a), the amplitude of the compressive DA solitary waves decreases with increase in $\theta_3$. While in Fig. 7(b) which rarefactive solitons are appeared, as $\theta_3$ increases, the amplitude of the DA solitary waves increases. Therefore, the large difference between the values of low and high ion temperature (leads to positive potential), appears in an increase of the DA solitary amplitude and the slight difference between the values of low and high ion temperature (leads negative potential), causes the solitons' amplitude to decrease.

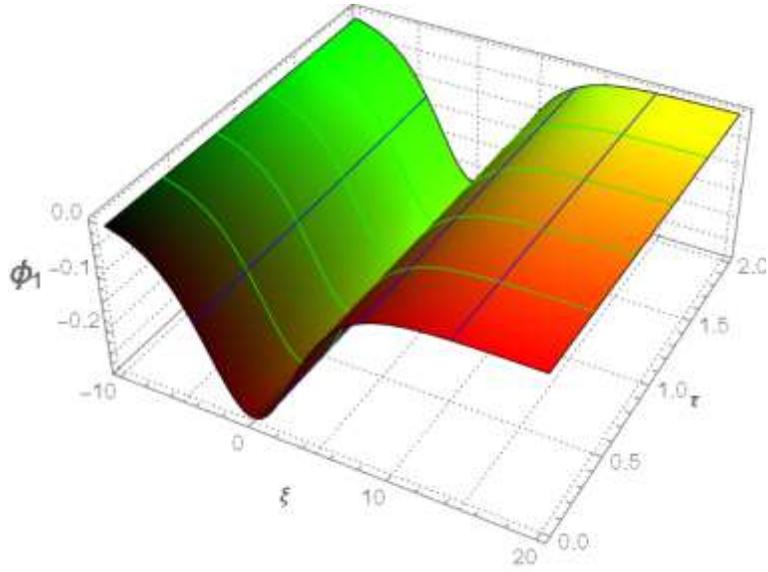

**FIGURE 1.** variation of $\phi_1$ position coordination $\xi$ and time $\tau$. We have chosen $\delta_1 = 0.3$, $\delta_2 = 0.7$, $\delta_3 = 5$, $\theta_1 = 0.2$ $\theta_2 = 0.1$, $\theta_3 = 0.5$ and $u_0 = 0.1$.

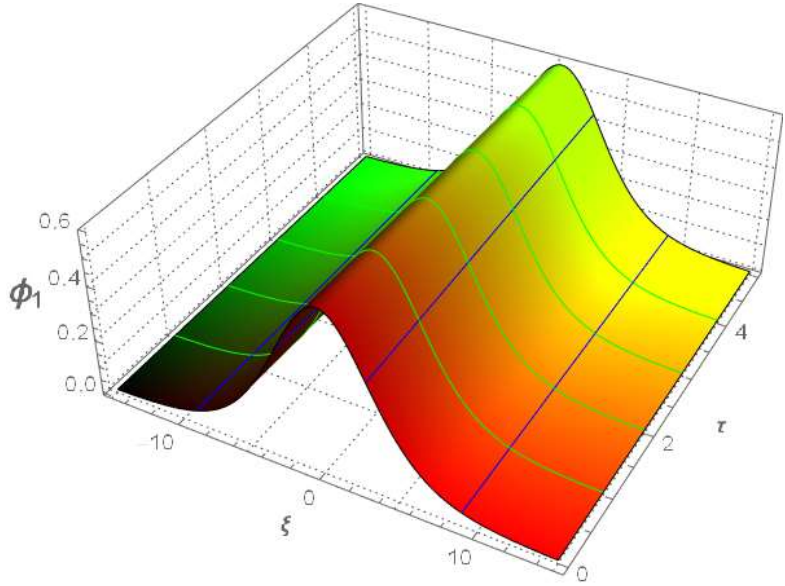

**FIGURE 2.** Variation of $\phi_1$ versus position coordination $\xi$ and time $\tau$. The calculations were performed by choosing $\delta_1 = 0.4$ $\delta_2 = 0.7$, $\delta_3 = 8$, $\theta_1 = 0.02$, $\theta_2 = 0.01$, $\theta_3 = 0.05$ and $u_0 = 0.1$.

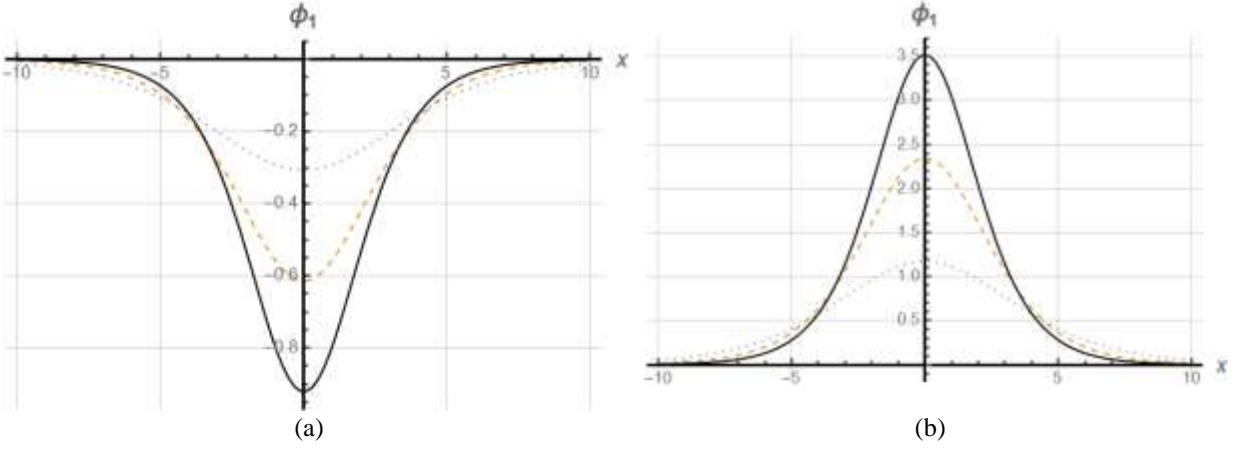

**FIGURE 3.** Variation of $\phi_1$ versus spatial coordinate $\chi$ for three different values of $u_0$. $u_0 = 0.1$ (dotted curve), $u_0 = 0.2$ (dashed curve), $u_0 = 0.3$ (solid curve). (a): negative potential for $\theta_1 = 0.2$, $\theta_2 = 0.1$ and $\theta_3 = 0.4$ (b): positive potential for $\theta_1 = 0.03$, $\theta_2 = 0.01$ and $\theta_3 = 0.05$. Other parameters are $\delta_1 = 0.2$, $\delta_2 = 0.3$, $\delta_3 = 5$.

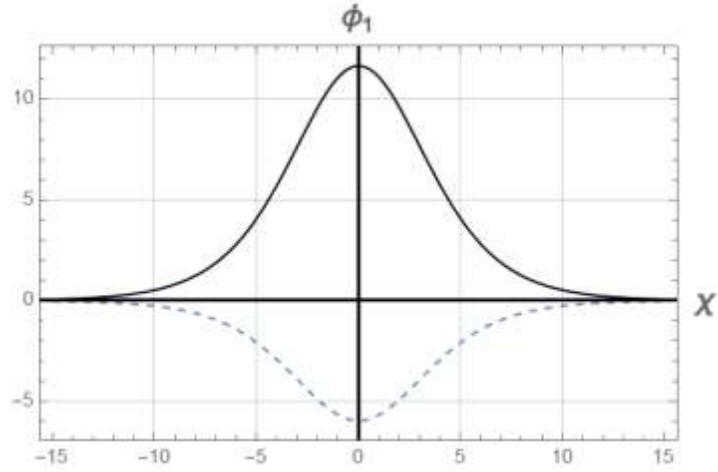

**FIGURE 4**. Variation of $\phi_1$ versus spatial coordinate $\chi$ for two different density ratio $\delta_3$, $\delta_3 = 6$ (dashed curve) and $\delta_3 = 7$ (solid curve). Other parameters are $\delta_1 = 0.2$, $\delta_2 = 0.6$, $\theta_1 = 0.02$, $\theta_2 = 0.01$, $\theta_3 = 0.08$ and $u_0 = 0.1$. ( $T_{eh} = 2T_{el} = 8T_{ih} = 100T_{il}$ )

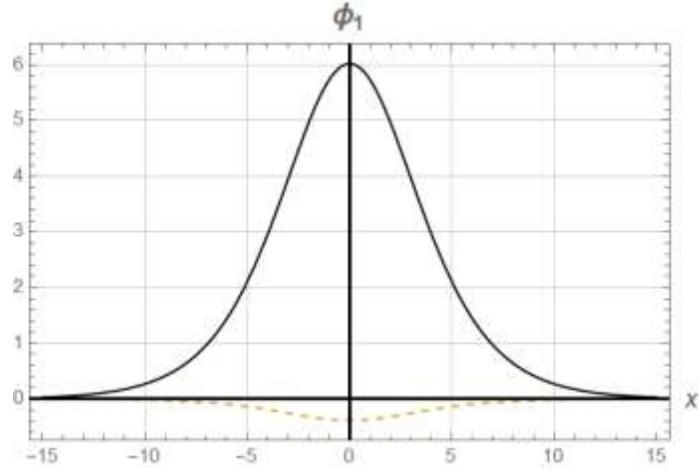

**FIGURE 5**. Variation of $\phi_1$ versus spatial coordinate $\chi$ for two different temperature ratio $\theta_3$ i.e. $\theta_3 = 0.2$ (dashed curve), $\theta_3 = 0.05$ (solid curve). The other parameters are $\delta_1 = 0.4$, $\delta_2 = 0.7$, $\delta_3 = 5$, $\theta_1 = 0.02$, $\theta_2 = 0.01$ and $u_0 = 0.1$.

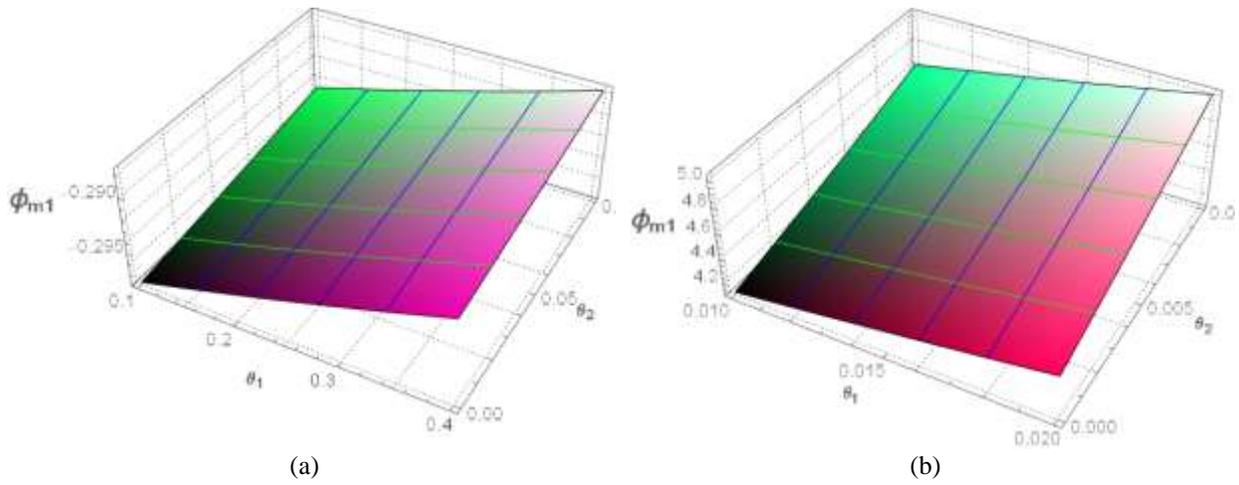

(a)  (b)

**FIGURE 6.** Variation of $\phi_{m1}$ versus $\theta_1$, $\theta_2$. (a): negative potential for $\theta_3 = 0.5$, (b): positive potential for $\theta_3 = 0.05$. The other parameters are $\delta_1 = 0.2$, $\delta_2 = 0.3$, $\delta_3 = 4$ and $u_0 = 0.1$.

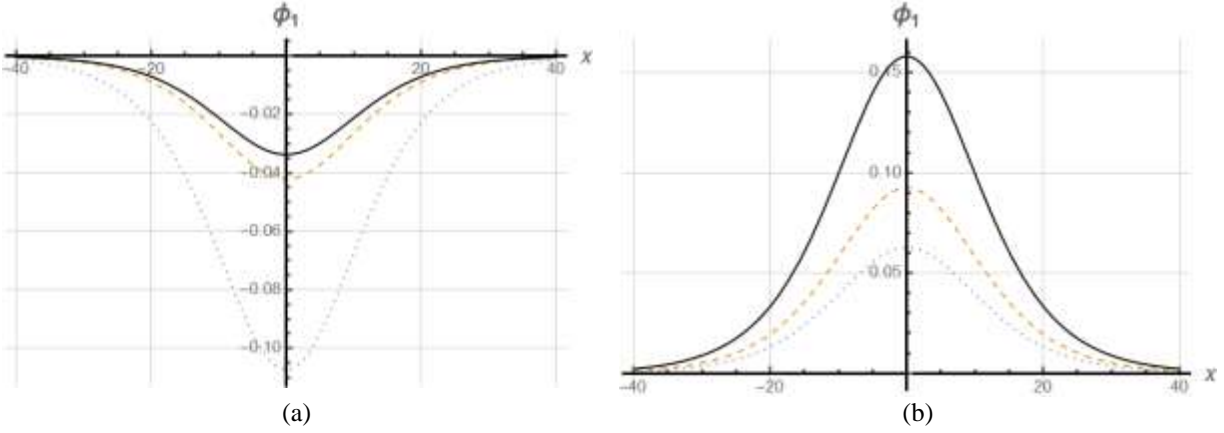

(a)                                                              (b)

**FIGURE 7.** Variation of $\phi_1$ versus spatial coordinate $\chi$ for different temperature ratios $\theta_3$. (a): negative potential for $\theta_3 = 0.1$ (dotted curve), $\theta_3 = 0.2$ (dashed curve), $\theta_3 = 0.3$ (solid curve), (b): positive potential for $\theta_3 = 0.03$ (dotted curve), $\theta_3 = 0.04$ (dashed curve), $\theta_3 = 0.05$ (solid curve). Other parameters are $\delta_1 = 0.3$, $\delta_2 = 0.4$, $\delta_3 = 5$ and $u_0 = 0.01$, $\theta_1 = 0.02$, $\theta_2 = 0.01$

## CONCLUSION

We have investigated the propagation of DASWs in an unmagnetized dusty plasma, consisting of Boltzmann distributed electrons with two distinct temperatures, negatively charged dust grains, and Boltzmann distributed ions with two different temperatures. Using the reductive perturbation technique, the K-dV equation is obtained. The results show that the present model supports both rarefactive and compressive DASWs structure. The presence of two different types of isothermal electrons and two different types of isothermal ions which appear in space plasma and laboratory environment, modify notably the basic features (polarity, amplitude, etc.) of solitary waves. It is found that in compressive DASWs polarity, the amplitude of the solitons increases with the increase of both $\theta_1$ (ratio of low temperature ion to low temperature electron) and $\theta_2$ (ratio of low temperature ion to high temperature electron) and in rarefactive DASWs polarity, as $\theta_1$ and $\theta_2$ increase, the amplitude of solitary waves increases.. Also, we found that the difference between the value of low and high ion temperature may be caused to change the DA soliton polarity. The positive potential solitary waves exist for lower values of the temperature ratio $\theta_3$ (ratio of low temperature to high temperature ions), while negative potentials appear for larger values of $\theta_3$. Furthermore, our results reveal that density ratio $\delta_3$ (ratio of the high temperature to low temperature ion densities) has effect on solitary wave structure and can change the polarity of DASWs.

# REFERENCES

1. C.K. Goertz, Rev. Geophys. 27, 271 (1989)
2. D.A. Mendis and M. Rosenberg, Annu. Rev. Astron. Astrophys. 32, 419 (1994)
3. P.K. Shukla and A.A. Mamun, Introduction to Dusty Plasma Physics, Institute of Physics, Bristol ( 2002)
4. O. Havens, F. Melandso, C.L. Hoz, T.K. Aslaksen, and T. Hartquist, Phys. Scr. 45, 491 (1992)
5. E.C. Whipple, T.G. Northrop, and D.A. Mendis, J. Geophys. Res. 90, 7405 (1985)
6. F. Verhesst, Astrophysics Space Sci. Library 30 (2000)
7. P.K. Shukla and D. Resendes, Phys. Plasmas 7, 1614 (2000)
8. V.E. Fortov, O.F. Petrov, V.I. Molotkov, M.Y. Poustylnik, V.M. Torchinsky, A.G. Khrapak, and A.V. Chernyshev, Phys. Rev. E 69, 016402 (2004)
9. J.B. Pieper and J. Goree, Phys. Rev. Lett. 77, 3137 (1996)
10. F. Verheest, V.V. Yaroshenko, G. Jacobs, and P. Meuris, Phys. Scr. 64, 494 (2001)
11. J. Winter, Physics of Plasmas. 7, 3862 (2000)
12. J. Cao and Themis Matsoukas, Journal of Applied Physics. 92, 2916 (2002)
13. R.L. Merlino, Physics of Plasmas. 16, 124501 (2009)
14. F. Verheest, Space Sci. Rev. 77, 267 (1996).
15. H. Kersten, G. Thieme, M. Fröhlich, D. Bojic, D. H. Tung, M. Quaas, H. Wulff, and R. Hippler, Complex (dusty) plasmas: Examples for applications and observation of magnetron-induced phenomena. Pure and applied chemistry. 77, 415 (2005)
16. T.M. Flanagan and J. Goree, Phys. Plasmas 17, 123702 (2010)
17. R.L. Merlino, A.J. Goree, Phys. Today 57, 32. (2004)
18. N.N. Rao, P.K. Shukla, M.Y. Yu, Planet. Space Sci. 38, 543 (1990)
19. A. Barkan, R.L. Merlino, N. D'Angelo, Phys. Plasmas. 2, 3563 (1995)
20. S. Paul, R. Denra, S. Sarkar, Eur. Phys. J. D. 74, 131 (2020)
21. R.E. Tolba, Eur. Phys. J. Plus. 136, 138 (2021)
22. A. Abdikian, S. Sultana, Physica Scripta. 96, 095602 (2021)
23. J. Tamang, B. Pradhan and A. Saha, IEEE Transactions on Plasma Science, 48, 3982 (2020)
24. S.K. El-Labany, W.F. El-Taibany, A.A. El-Tantawy, N.A. Zedan, contributions to plasma physics. 60, 10 (2020)
25. P.F. Li, C.R Du, Physics Letters A. 384, 27 (2020)
26. A. Davletov, F. Kurbanov, Y. Mukhametkarimov and L. Yerimbetova, IEEE Transactions on Plasma Science. 49, 2007 (2021)
27. Y. Saitou, Physics of Plasmas. 28, 073703 (2021)
28. Y. Nakamura and H. Sugai, Chaos Solitons Fractals. 7, 1023 (1996)
29. G. Gloeckler, J. Geiss, H. Balsiger, P. Bedini, J.C. Cain, J. Fischer, L. Fisk, A. Galvin, F. Gliem, and D. Hamilton, Astron. Astrophys. Suppl. Ser. 92, 267 (1992)
30. B. Buti, Phys. Lett. A. 76, 251 (1980)
31. Y. Nishida, T. Nagasawa. Phys. Fluids. 29, 345 (1986)
32. A. Shome, S. Pramanik, contributions to plasma physics. 61, 7 (2021)
33. K. Annou, Astrophys Space Sci 350, 211 (2014)
34. M. Emamuddin, S. Yasmin, M. Asaduzzaman, and A.A. Mamun, Phys. Plasmas 20, 083708 (2013)
35. M.G.M. Anowar and A.A. Mamun, IEEE Transactions on Plasma Science. 37, 1638 (2009)
36. X. ZHong, H. Chen, N. Liu and S. Liu, Pramana J. Phys. 86, 885 (2016)
37. A. Varghese, A.C. Saritha, N.T. Willington, M. Michael, S. Sebastian, G. Sreekala and C. Venugopal J.Astrophys.Astr. 41,1 (2020)
38. S. K. Maharaj, R. Bharuthram, S.V. Singh, and G.S. Lakhina Phys. Plasmas. 19, 122301 (2012)
39. M. Shahmansouri and H. Alinejad Phys. Plasmas. 20, 082130 (2013)
40. F. Araghi, D. Dorranian, Plasma Physics Reports. 42, 155 (2016)
41. M. Asaduzzaman, A.A. Mamun, Astrophys Space Sci. 341, 535 (2012)
42. M.N. Alam, A.R. Seadawy, D. Baleanu, Alexandria Engineering Journal. 59, 1505 (2020)
43. D.N. Gao, J.B. Yue, J.P. Wu et al. Plasma Phys. Rep. 47, 48(2021)
44. R. Srivastava, H.K. Malik, D. Singh, J Theor Appl Phys 14, 11(2020)
45. M.S.Alam, M.G.Hafez, M.R. Talukder, and M. Hossain Ali, Physics of Plasmas. 24, 103705 (2017)

46. S.G. Tagare, Physics of Plasmas 4, 3167 (1997)
47. S.K. El-Labany, E.F. El-Shamy, M.Shokry, physics of plasma. 17, 113706 (2010)
48. J. Borhanian, M. Shahmansouri, Physics of Plasmas. 20, 013707 (2013)
49. X. Bai Song, H.K. Fen, H.Z. Qia, Chin.phys.Lett. 17, 815 (2000)
50. D. Dorranian, A. Sabetkar, Physics of Plasmas. 19, 013702 (2012)
51. H.R. Pakzad, K. Javidan, Chaos, Solitons and Fractals. 42, 2904 (2009)
52. T.S. Gill, N.S Saini, H. Kaur, Chaos, Solitons and Fractals. 28 1106 (2006)
53. K. He, H. Chen, S. Liu. Jpn. J.Appl. Phys. 59, 016001 (2020)
54. A.Mannan, S.D. Nicola,R.Fedele, A.A. Mamun, AIP Advances. 11, 025002 (2021)
55. L. Kavitha, K. Raghavi, C. Lavanya, M. Kailas, D. Gopi, Transactions on Plasma Science, 49, 546 (2020)
56. Z. Rezaiinia, T. Mohsenpour, S. Mirzanejhad, Contrib. Plasma Phys. 59, 252 (2019)
57. H.K. Malik, R. Srivastava, S. Kumar and D. Singh, Journal of Taibah University for Science. 14, 417 (2020)
58. M. Emamuddin, M.M. Masud, A.A. Mamun, Astrophys Space Sci 349, 821 (2014)
59. S.K. El-Labany, E.F. El-Shamy, W.F. El-Taibany, W.M. Moslem, Chaos, Solitons and Fractals. 34 1393 (2007)
60. H. Washimi and T. Taniuti, Phys. Rev. Lett. 17, 996 (1966)
61. A.A. Mamun, R.A. Cairns, and P.K. Shukla, Phys. Plasmas 3, 702 (1996)